\documentclass[aps,prb,superscriptaddress,twocolumn,showpacs,amsmath,amssymb,amsfonts,floatfix]{revtex4-1}
\usepackage{graphicx,color,dcolumn,bm}
\usepackage{CJK}
\usepackage{times}
\usepackage{bbm}
\usepackage[colorlinks=true,citecolor=blue,urlcolor=blue]{hyperref}
\usepackage{natbib}

\begin{document}
\begin{CJK}{UTF8}{bsmi}
\title{Detection of symmetry-protected topological phases in one dimension with \\multiscale entanglement renormalization}
\author{Hsueh-Wen Chang(張學文)}
\affiliation{Center for Theoretical Science and Department of Physics, National Taiwan University, Taipei 106, Taiwan}

\author{Yun-Da Hsieh (謝昀達)}
\affiliation{Center for Theoretical Science and Department of Physics, National Taiwan University, Taipei 106, Taiwan}

\author{ Ying-Jer Kao (高英哲)}
\affiliation{Center for Theoretical Science and Department of Physics, National Taiwan University, Taipei 106, Taiwan}
\affiliation{Center for  Advanced Study in Theoretical Science, National Taiwan University, Taipei 10607, Taiwan}
\date{\today}

\begin{abstract}
Symmetry-protected topological (SPT) phases are short-range entangled quantum phases with symmetry, which have  gapped excitations in the bulk and gapless  modes at the edge.  In this paper, we study the SPT phases in the spin-1 Heisenberg chain with a single-ion anisotropy $D$, which has a quantum phase transition between a Haldane phase and a large-$D$ phase. Using symmetric  multiscale entanglement renormalization ansatz (MERA) tensor networks, we study the nonlocal  order parameters for time-reversal and inversion symmetry.  For the inversion symmetric MERA, we  propose a brick-and-rope representation  that gives a  geometrical interpretation of inversion symmetric tensors. Finally, we  study the symmetric renormalization group (RG) flow of  the inversion symmetric string-order parameter, and show that entanglement renormalization with symmetric tensors produces  proper behavior of the RG fixed-points.
 \end{abstract}

\maketitle
\end{CJK}

\section{Introduction}
Phases of matter are traditionally characterized  by local order parameters associated with spontaneous symmetry breaking. This is  the essence of the Landau paradigm of phase transitions. However, there are states of matter which fall beyond this type of characterization. 
One  famous example is the fractional quantum Hall state, which is topologically ordered with no local order parameter.\cite{Wen:1990fk}  Topological phases are often characterized by a gapped  ground state in the bulk and the presence of gapless edge modes.\cite{Wen:1995kx}  Classification of these topological phases remains difficult since there is no-symmetry breaking associated with these phases.  Using the ideas based on local unitary transformations, it is argued 
that an \textit{intrinsic} topological order is associated with the
pattern of long-range entanglements in gapped quantum
systems with finite topological entanglement entropy.\cite{Chen:2010fk} 

The situation in one dimension(1D) is quite different. The \textit{intrinsic} 
topological order as described above exists only in two and higher dimensions since  
the ground states of all  one-dimensional  gapped spin systems are in a single phase.\cite{Chen:2011vn} However, if we impose symmetries on the system, there might exists topological phases which are protected by the presence of the symmetry. These symmetry-protected topological (SPT) phases are formed by gapped short-range-entangled quantum states
that do not break any symmetries.\cite{Gu_Wen2009}
The Haldane phase of odd-integer spin chains\cite{Haldane1,*Haldane2} is an example of a SPT phase in one dimension.\cite{Gu_Wen2009,Pollmann2010}
In addition, by tuning  some parameters in the Hamiltonian, it is possible to drive a quantum phase transition between two topologically inequivalent phases through a  critical point. Classification of these SPT phases is not easy since there exists no spontaneous symmetry breaking, i.e., no local order parameters,  on both sides of the critical point. It is desirable to find  some quantities that change through the critical point to characterize the phase transition.\cite{Pollmann2010}

One representative example of the Haldane phase is the spin-1 AKLT state, which is the ground state of an exact solvable  AKLT model.\cite{AKLT} The AKLT state exhibits no conventional  long-range order, but can be characterized by nonlocal string-order and edge states.\cite{Nijs:1989fk,Kennedy:1992vn}
By measuring a string of identical operators with distinct end points, it can be shown the system exhibits correlations that are independent of the string length, indicating a string-order in the system. Later, it is realized that the string-order parameter can be generalized to distinguish different SPT phases, if  symmetry operations on the entanglement between blocks is considered.\cite{Pollmann2010, Chen:2010fk} Studying the symmetry operation on the matrix product state (MPS), nonlocal string-order parameters have been constructed to characterized SPT states for different types of global symmetries.\cite{Pollmann2012,Haegeman:2012uq, Schuch:2011fk} In this work, we demonstrate how to use a symmetry-protected  multiscale entanglement renormalization ansatz (MERA)\cite{Singh:2010zr,Singh:2013ly} to calculate string-order parameters in the presence of time-reversal and  inversion symmetry  for a  spin-1 chain.

MERA is a real-space renormalization group (RG) method based on the concept of entanglement renormalization.\cite{Vidal2007,MERA_algorithm} By inserting disentanglers into the system to remove short-range entanglement before renormalization, it prevents the accumulation of degrees of freedom during the real-space RG transformation.  This method have been applied to several 1D and 2D systems,\cite{MERA_algorithm,Evenbly:2009bh,Evenbly:2010qf,Evenbly:2010dq} and even at quantum critical points.\cite{Pfeifer:2009ve}
MERA with a global ${Z}_2$ symmetry, which the time-reversal symmetry belongs to, has been proposed for the case of  parity.\cite{MERA_fermion} On the other hand,  the algorithm for constructing an inversion-symmetric MERA is not yet available in the literature.\cite{Evenbly:fk} 
Examining the coarse-graining process of MERA and the relation between the geometrical inversion and the tensor symmetry, we construct an inversion symmetric MERA based on a ``brick-and-rope'' representation. This enables us to study  the inversion symmetric string-order parameter using inversion symmetric MERA. 

This paper is structured as follows: We first introduce the spin-1 chain model and review some of its properties in Sec.~\ref{sec:Model}. In  Sec.~\ref{sec:basicMERA}, we discuss how to implement a ${Z}_2$ symmetric MERA and introduce our algorithm on  inversion symmetric MERA based on a brick-and-rope representation. In Sec.~\ref{sec:StringOrder}, we present the results of string-order parameters using both time-reversal and inversion symmetric MERA. We conclude our work in Sec.~\ref{sec:Conclusion}.

\section{Model}
\label{sec:Model}
We study the spin-1 Heisenberg model with a  single-ion anisotropy term on a periodic chain of length $N$:
\begin{equation}
H = \sum_{i=1}^N \textbf{S}_i \cdot \textbf{S}_{i+1} + D \sum_{i=1}^N (S^z_i)^2.
\label{Hamiltonian}
\end{equation}
For $D\rightarrow 0$, the ground state is in the Haldane phase, which is topologically non-trivial.\cite{Gu_Wen2009}  The AKLT state, which is exact solvable, is known to describe the same phase as the Haldane phase. \cite{AKLT} For $D \gg 1$, the ground state is in the large-$D$ phase,\cite{Gu_Wen2009} which is topologically equivalent to the product state  $|0\rangle^{\otimes N}$ in the $S_z$ basis and is a topologically trivial state. By tuning $D$, there is a Gaussian transition from the Haldane phase to the large-$D$ phase near $D_c  = 0.96845(8)$.\cite{Hu:2011lr} This model has translation, spatial inversion, time-reversal symmetries, and a $Z_2\times Z_2$  symmetry of rotations $\mathcal{R}_x=\exp(i\pi S^x)$ and $\mathcal{R}_z=\exp(i\pi S^z)$.\cite{Kennedy:1992vn,Nijs:1989fk} For the purpose of this paper, we will focus on inversion and  TR symmetries, while the translation symmetry is considered explicitly in the MERA tensor network.

\section{Symmetries in Multiscale-Entanglement Renormalization Ansatz}
\label{sec:basicMERA}

 In this section, we will briefly review the algorithms for MERA, and discuss how to incorporate the  time-reversal (${Z}_2$) symmetry.\cite{MERA_algorithm,MERA_fermion} Finally, we will give  a detailed description of  an implementation for the inversion symmetry.   
 
\subsection{Multiscale-Entanglement Renormalization Ansatz}

\begin{figure}
\includegraphics[width=0.9\columnwidth]{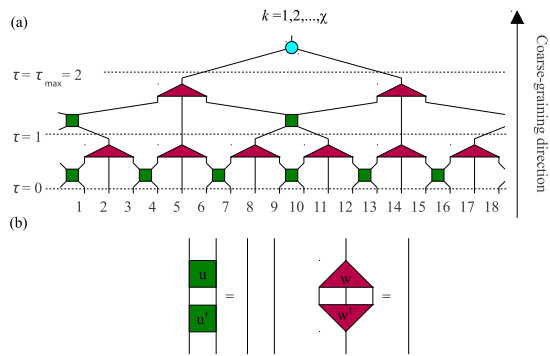}
\caption{(Color online) (a) The structure of  an 18-site 1D ternary  MERA with the periodic boundary condition.  In addition to the (horizontal) physical  direction, there is also  a (vertical) coarse-graining direction. While coarse-graining, we truncate the less important degree of freedom by $w$. At the top layer,  a reduced basis of $\chi$ states is kept and  we find the minimum energy state inside this truncated Hilbert space. (b) Diagrammatic representation of the isometric constraints  for the disentanglers $u$ and the isometries $w$.
}\label{MERA_UW}
\end{figure}

\begin{figure}
\includegraphics[width=0.9\columnwidth]{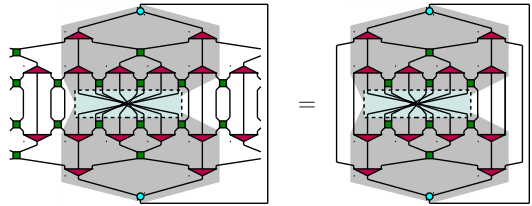}
\caption{(Color online) Diagram of an inversion symmetric string-order parameter of eight inverted sites in MERA. Shaded area corresponds to the causal cone in which the tensor contraction needs to be performed. The width of the causal cone grows with the size of the swapped region. }
\label{inverSO_MERAex}
\end{figure}

 Figure~\ref{MERA_UW} shows an example of a 1D ternary (three-to-one)  MERA network. The fundamental tensors in the structure  are the disentangler ($u$), and the  isometry ($w$). From a RG transformation perspective, an isometry coarse-grains, in this case, three sites into one effective site. A disentangler $u$ acts across the boundary of two blocks of sites to remove short-range entanglement between the blocks.\cite{Vidal2007, MERA_algorithm} The $u$'s and $w$'s  satisfy the unitarity constraints (Fig.\ref{MERA_UW}):
\begin{align}
&{u}^{\dagger}{u} = {u}{u}^{\dagger} = \mathbb{I}^{\otimes 2}
\label{u_property}\\
&w^{\dagger}w = \mathbb{I}
\label{w_property}
\end{align}
 An important implication of these constraints is that the tensor contraction process can be simplified and only tensors inside a ``causal cone'' structure need to be contracted (shaded area in Fig.~\ref{inverSO_MERAex}). 
To facilitate our discussion on symmetry operations in the following, we will interpret $u$ as a transformation of the incoming wave function in the \textit{original} basis, while $w$ corresponds to  a transformation to a \textit{truncated} basis with $\chi$ basis states.

Since our model is translationally invariant, we set the disentanglers and isometries in each layer identical. In the simulation, 
we initiate the tensors obeying the constraints Eqs.~(\ref{u_property}) and (\ref{w_property}), and  perform updates  respecting these conditions. We treat the tensor coefficients  as variational parameters, and optimize $u$'s and $w$'s by minimizing the energy.\cite{MERA_algorithm} 
We perform updates on $u$ or $w$ by a  singular value decomposition (SVD) of the environment tensor for the given tensor. We refer interested readers to Ref.~\onlinecite{MERA_algorithm} for details of this procedure.
Figure~\ref{inverSO_MERAex} shows a typical calculation of the inversion symmetric string-order parameter. The width of the causal cone scales with the number of the inverted sites, which must be contained in the causal cone at the physical layer.  To simplify the contraction, we  want to use symmetric tensors  to  remove these symmetry-related operators.

\subsection{${Z}_2$ symmetric MERA}

Several important symmetry operations, such as  time-reversal (TR) and parity fall into a general class of an internal ${Z}_2$ symmetry. In the following, we will use the TR symmetry  for spin-1 as a concrete example to discuss the algorithm, and it can be easily generalized  to any internal ${Z}_2$ symmetries.\cite{MERA_fermion}

\begin{figure}
\includegraphics[width=\columnwidth]{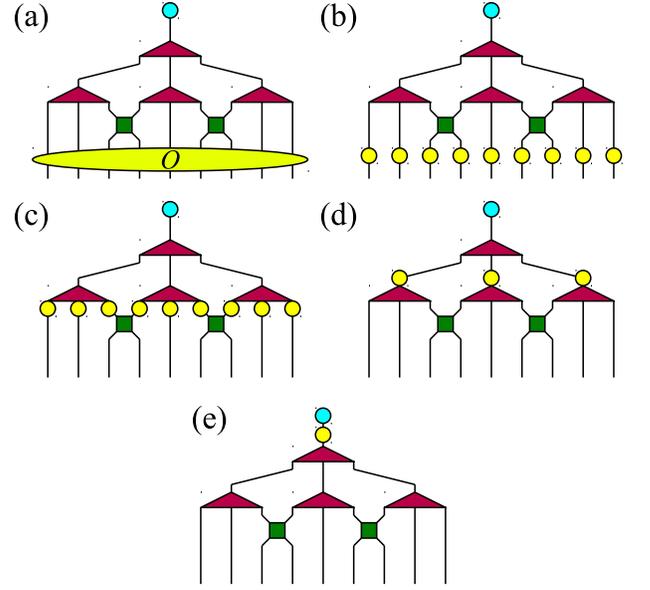}
\caption{ (Color online) Commuting a TR symmetric MERA with  TR operators (yellow circles).   We want the operator in (a) to "jump" to the top as in (e). By enforcing all the $w$'s and $u$'s to be TR symmetric, we can  break the TR operator into a product of local operators as shown in (b). Therefore, each operator can commute with the tensors locally, and we can \textit{lift} the operators to the top of MERA through steps shown in (c)-(e).
}\label{TRonMERA}
\end{figure}

Define the global TR symmetry operator $O\equiv (O_T)^{\otimes  N}$,
where $O_T$ is the TR symmetry operator acting on each bond with $O_T^2=1$. To represent a symmetric state, we use a MERA composed of symmetric tensors,\cite{Singh:2010zr} where the symmetric isometry, $w$, and disentangler, $u$, satisfy
\begin{align}
&(O_T \otimes O_T)u^{\tau}(O_T^{\dagger} \otimes O_T^{\dagger}) = u^{\tau}\\
&(O_T )w^{\tau}(O_T^{\dagger} \otimes O_T^{\dagger} \otimes O_T^{\dagger}) = w^{\tau}
\end{align}
With these conditions, the TR symmetry is preserved during the RG transformation and  the Hamiltonian at the $\tau$-th layer, $H^{\tau}$, is also TR symmetric.  We obtain the ground state in a specific symmetry sector by diagonalizing the top Hamiltonian $H^{\tau_{\rm max}}$. The top density matrix is also TR symmetric and the symmetry is also preserved as we descend down to the physical layer.
 Treating the $u$ and $w$ tensors as linear operators, the MERA network, denoted by $\mathcal{M}$, can be regarded as a collection of operators with internal bonds contracted.   Therefore, a TR symmetric MERA should also satisfy the operator identity
\begin{equation}
\label{eq:commute}
[\mathcal{M},O] = 0.
\end{equation}
Graphically, Eq.~\ref{eq:commute} can be represented by Figs.~(\ref{TRonMERA}a) and (e). Since the TR symmetry is an internal symmetry, we can decompose the global operator $O$ into local symmetry operators $O_T$ as in Fig.~(\ref{TRonMERA}b). If each tensor commutes with the local TR operator $O_T$, we can \textit{lift} the TR operators layer by layer to obtain Fig.~(\ref{TRonMERA}e).  In terms of the tensor elements, these TR-symmetric tensors are block-diagonal in  a TR-symmetric basis. 
We now explain how to build TR-symmetric bases as we coarse grain the lattice via MERA.  
\begin{figure}
\includegraphics[width=\columnwidth]{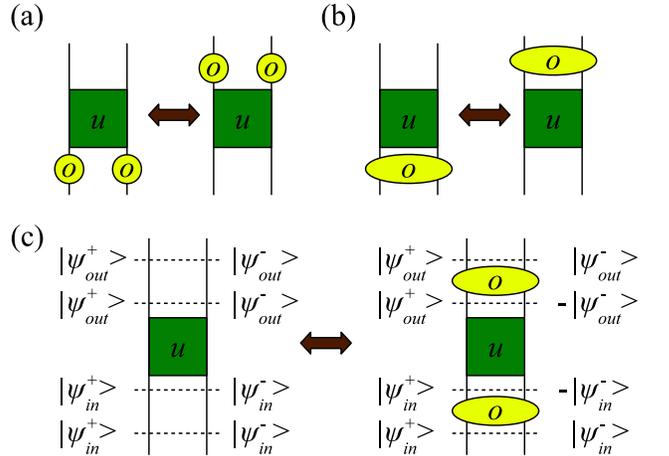}
\caption{ (Color online) (a) Commuting two local TR operators with a disentangler $u$. (b) A global TR operator is constructed by grouping two local TR operators.  (c) Applying another global TR operator at the bottom, in the diagram at the left, the operator becomes an identity while at the right,   there are two TR operators above and below $u$.  (Left) For a TR symmetric $u$, the in-state and out-state should be in the same symmetry sector. (Right)  Application of the TR operator transforms the wave function as $|\psi_{in}^{\pm}\rangle \rightarrow \pm |\psi_{in}^{\pm}\rangle$. After the action of $u$, the states are still TR symmetric.  Finally, after the second TR operator, we obtain the same out-states as those at the left. 
}\label{TRonUW}
\end{figure}


For a MERA, or any other tensor network, to have a ${Z}_2$ symmetry, we require the tensors ($w$ and $u$) to preserve the ${Z}_2$ symmetry; that is, 
\begin{equation}
T_{i_1 i_2 \ldots i_M} = 0 \mbox{, if } Z(i_1)Z(i_2)\cdots Z(i_M) \neq 1;
\label{tensor_Z2sym}
\end{equation}
where $Z(i_k) = \pm 1$, and the sign is $+ (-)$ if the state labeled by $i_k$ is ${Z}_2$ symmetric even (odd). 
To transform the basis into a ${Z}_2$ symmetric basis in the upper layers, we treat a tensor as an operator on a state by defining the \textit{in}- and \textit{out}-directions.   The tensor is symmetric if the \textit{in}- and \textit{out}-states belong to the same symmetry sector (Fig.~({\ref{TRonUW}c)).
To construct a $Z_2$ symmetric MERA, we first start from the physical layer and go up (Fig.~\ref{MERA_UW}). We go \textit{in} the tensor from below and go \textit{out} from the top. We first encounter the disentangler $u$, and the in-state is a linear combination of two physical sites. The symmetry of a combined state obeys the fusion rules:
\begin{equation}
\begin{aligned}
&(+) \otimes (+) \rightarrow (+), \; (-) \otimes (-) \rightarrow (+)\\
&(+) \otimes (-) \rightarrow (-), \; (-) \otimes (+) \rightarrow (-),
\end{aligned}
\label{parity_fusion}
\end{equation}
that is, if  two sites are both in a ${Z}_2$-symmetry even state, the combined state is  ${Z}_2$-symmetry even, etc. Recall that the operation of $u$ can be regarded as a unitary transformation of the wave function in the original basis;  therefore, if $u$ preserves the symmetry, the tensor elements of $u$ corresponding to a transform from the in-states to the out-states in different symmetry sectors must be zero, which is exactly   Eq.~(\ref{tensor_Z2sym}). 
Next comes an isometry $w$ which coarse-grains three sites into one. If there is no truncation, we can regard $w$ as a unitary transformation of the basis; i. e.,  the elements of $w$ should give us information about how the new basis is formed by a linear combination of the incoming basis. To keep the outgoing basis ${Z}_2$ symmetric,  we have to make sure that we do not mix  bases belonging to different symmetry sectors. We perform truncations on the resulting $Z_2$ symmetry even and odd basis states independently. 
This process is repeated layer by layer to construct the $Z_2$-symmetric MERA.

In practice, we  initiate the tensors obeying the unitary constraints Eqs.~(\ref{u_property}) and  (\ref{w_property}), together with  Eq.~(\ref{tensor_Z2sym}). During the SVD updates, the first two conditions are automatically satisfied,\cite{MERA_algorithm} and  the $Z_2$ symmetry is also satisfied because the  environment  satisfies Eq.~(\ref{tensor_Z2sym}), as  it is formed by ${Z}_2$-symmetric tensors. Therefore, in the symmetric basis (of the total in  and out states), the environment tensor should be block-diagonal. Therefore we can perform  the SVD  block by block, and the updated tensor will also preserve the symmetry. For the same reason, the Hamiltonian at each layer will also be symmetric since the ascending operator is again obtained by contracting symmetric tensors. Finally, the top Hamiltonian is also block-diagonal. We can diagonalize the block  for a specific symmetry sector to find the ground state. 
\begin{figure}[tbp]
\includegraphics[width=0.9\columnwidth]{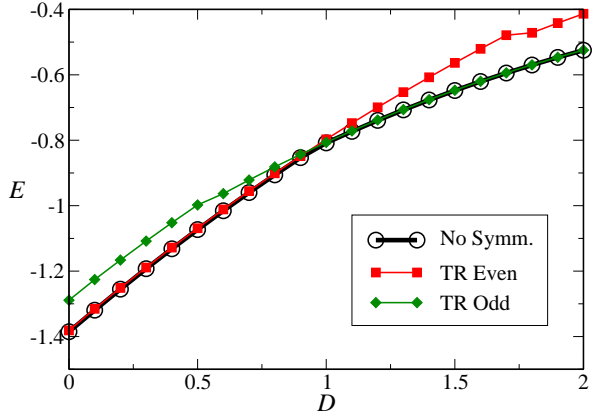}
\caption{ (Color online) Lowest energy states obtained by MERA with (green diamond and red square) or without (black circle) imposing the time-reversal symmetry for $N = 9, \chi = 12$ ($\chi_{\rm even}=\chi_{\rm odd}=\chi/2$). 
}\label{Data_E_TR}
\end{figure}

To benchmark the algorithm, we show the ground states obtained by the TR symmetric MERA in the TR even and odd  sectors, and compare it with the results from the non-symmetric MERA algorithm in Fig.~\ref{Data_E_TR}. We show the results for an $N=9$ chain with $\chi = 12$ ($\chi_{\rm even}=\chi_{\rm odd}=\chi/2$). The small system size is chosen such that the energy splitting between the states in different symmetry sectors is large deep inside the phase. For large $D$, the ground state is a product state  $|0\rangle^{\otimes N}$. Under TR, $|0\rangle \rightarrow -|0\rangle$,  and the ground state is TR odd. For small $D$, the ground state is  the AKLT state, which is TR even.\cite{AKLT} This is a clear level crossing close to $D=1$,  indicating a phase transition.

Before moving to the next subsection, we would like to discuss how a ${Z}_2$ symmetric MERA can simplify the calculations by \textit{removing}  the ${Z}_2$ operator (denoted as $O_{{Z}_2}$) in the calculation. 
Consider a $Z_2$ operator $(O_{{Z}_2})^{i_1}_{i_1^{\prime}}$ that operates on bond $i_1^{\prime}$ of a symmetric disentangler $u^{i_1^{\prime}i_2}_{i_3i_4}$. Assuming that the bond $i_1^{\prime}$ is ${Z}_2$ symmetry even for $i_1^{\prime} = 1,2,\ldots,\chi_{\rm even}$ and odd for the rest $\chi_{\rm odd}=\chi-\chi_{\rm even}$. The tensor element with $i_1^{\prime}$ index for ${Z}_2$ symmetry even will not change, while the odd ones are multiplied by $-1$ after the application of $O_{{Z}_2}$. In our example,  a TR operator operates on a TR symmetry even basis does nothing, while the operation change the sign of the state if the state is TR symmetry odd. That is, in a ${Z}_2$ symmetric basis:
\begin{equation}
(O_{{Z}_2})^i_{i^{\prime}} = \delta_{ii^{\prime}}Z(i).
\end{equation}
 Therefore, for a general ${Z}_2$ symmetric tensor $T$, we have:
\begin{align}
&(O_{{Z}_2})^{i_1}_{i_1^{\prime}}\cdots(O_{{Z}_2})^{i_k}_{i_k^{\prime}} T_{i_1^{\prime}\dots i_k^{\prime}i_{k+1} \ldots i_M}\nonumber\\
&= Z(i_1)\cdots Z(i_k) T_{i_1\ldots i_M}, k \leq M
\label{Z2operation}
\end{align}
Now if $k = M$ in Eq.~(\ref{Z2operation}), combining with Eq.~ (\ref{tensor_Z2sym}), we have:
\begin{align}
&(O_{{Z}_2})^{i_1}_{i_1^{\prime}}\cdots(O_{{Z}_2})^{i_M}_{i_M^{\prime}} T_{i_1^{\prime}\dots i_M^{\prime}}\nonumber\\
&= Z(i_1)\cdots Z(i_M) T_{i_1\ldots i_M} = T_{i_1\ldots i_M}
\label{Z2sym_tensor}
\end{align}
since $T_{i_1\ldots i_M} \neq 0$ only if $Z(i_1)\cdots Z(i_M) = 1$. Therefore, if we operate $O_{{Z}_2}$ on all the bonds of a ${Z}_2$- symmetric tensor, the operations cancels each other. In addition, since $(O_{{Z}_2})^2 = \mathbb{I}$, we have (see Fig.~\ref{jump_move}):
\begin{equation}
\begin{aligned}
(O_{{Z}_2})^{i_1}_{i_1^{\prime}}\cdots(O_{{Z}_2})^{i_k}_{i_k^{\prime}} T_{i_1^{\prime}\ldots i_k^{\prime}i_{k+1} \ldots i_M}\\
=(O_{{Z}_2})^{i_{k+1}}_{i_{k+1}^{\prime}}\cdots (O_{{Z}_2})^{i_M}_{i_M^{\prime}} T_{i_1\ldots i_k i_{k+1}^{\prime} \ldots i_M^{\prime}}
\end{aligned}
\label{eq_jump_move}
\end{equation}
This is essentially the \textit{jump move}  described in Ref.~\onlinecite{MERA_fermion}. We can   move the symmetry operators to the boundary of the causal cone, and maintain the  simple causal cone structure of a multi-site correlator in the calculation of these string-order parameters. 
\begin{figure}[tbp]
\includegraphics[width=0.9\columnwidth]{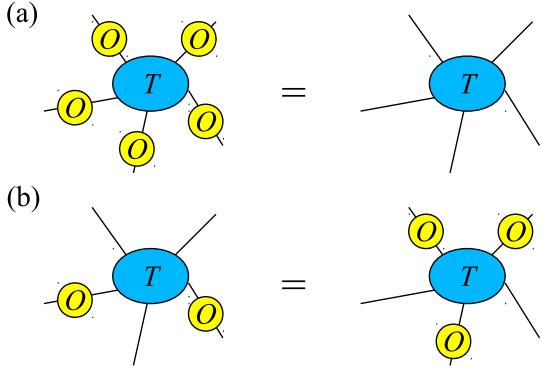}
\caption{ (Color online) Diagrammatic representations of (a) Eq.~(\ref{Z2sym_tensor}) and (b) Eq.~(\ref{eq_jump_move}). $T$ is a general symmetric tensor, and $O$ is a $Z_2$ symmetry operator. The result of symmetry operations on some bonds of a symmetric tensor is the equivalent to the case where the operators operate on all the other bonds. 
}\label{jump_move}
\end{figure}

\begin{figure}[th]
\includegraphics[width=0.9\columnwidth]{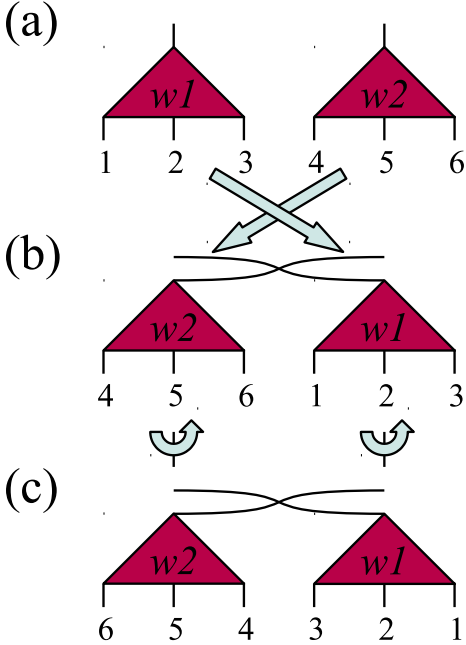}
\caption{(Color online)  Geometrically, it takes two steps to invert the state composed of the product of two  $w$'s (a).  First, we swap the position of the $w$'s to obtain (b). Then we invert the $w$'s about their own central bonds to obtain (c). Finally,  six sites are inverted.
}\label{inver2w}
\end{figure}

\subsection{Inversion symmetric MERA}\label{inversion_MERA}
Incorporating   inversion symmetry into MERA is non-trivial since the spatial inversion is defined  \textit{globally}. On the other hand, the tensors are \textit{local} in the sense that they are only connected to part of the network. To perform a global  inversion with local tensors requires special care. Figure~\ref{inver2w} shows an example of how to geometrically invert a state composed of the product of two $w$'s. We first swap the top bonds for each tensor, and then invert each $w$ about its own central bond. Hence, we can obtain a global inverted state by locally invert the tensors one-by-one geometrically. To understand how this geometrical operation can be translated into local tensor operations, we need to construct inversion symmetric basis by symmetrize or anti-symmetrize the original incoming and outgoing basis according to the fusion rules (see Appendix~A).  When the in- and out-basis are both transformed into the global inversion symmetric basis, an  inversion symmetric tensor  is block-diagonal. This is an analogy of Eq.~(\ref{tensor_Z2sym}) for  inversion. In this case, the inverted state will be either the state itself (inversion even) or gains a minus sign (inversion odd). 

\begin{figure}[tb]
\includegraphics[width=0.9\columnwidth]{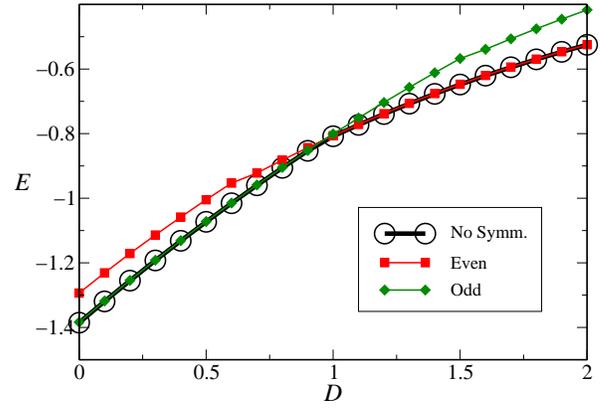}
\caption{(Color online) Lowest energy states obtained by MERA with (green diamond and red square) or without (black circle) imposing the inversion symmetry for $N = 9, \chi = 12 (\chi_{\rm even}=\chi_{\rm odd}=\chi/2)$. 
}\label{Data_E_inver}
\end{figure}

Figure~\ref{Data_E_inver} shows the result for an $N=9$ chain with $\chi = 12$ ($\chi_{\rm even}=\chi_{\rm odd}=\chi/2$). The AKLT phase is known to be inversion odd for a periodic chain with odd number of sites\cite{AKLT}, and the large-$D$ phase is inversion even.  A level crossing  near $D=1$ indicates a topological phase transition from the topologically nontrivial Haldane phase to the topologically 
trivial large-$D$ phase. 

Before ending this subsection, we would like to discuss how one can simplify calculations  in an inversion symmetric MERA by removing the inversion symmetry operator. This is more difficult than in the ${Z}_2$-symmetry case, since the inversion operator can not be decomposed into local operators acting only on a single bond. Here,  we present an analogy 
  between the inversion of symmetric tensors  and the geometrical inversion,  called the \textit{brick-and-rope} representation, to help us construct the necessary symmetry operations for inversion symmetric tensors.
  
Imagine one constructs a real-life MERA tensor network  with \textit{bricks}, representing tensors,   and \textit{ropes}, representing bonds (Fig.~\ref{inverMERA}(a)). To obtain the inverted state,  one needs only to rotate the whole structure by $180^\circ$ (Fig.~\ref{inverMERA}(b)). After the rotation, the structure would look very similar to the original except two crucial differences: the bricks are now turned to their \textit{backsides}, and the top rope is \textit{twisted}. 
Intuitively, given an inversion symmetric tensor, it is natural to argue that due to the inversion symmetry,  the``backside'' of the tensor should be the same as the tensor itself .  Also, since the inverted MERA tensor network should represent the inversion of a state, a ``twisted rope'' would thus represent an inverted state. In the following, we will discuss how these physical intuitions can be realized using local tensor operations.

\begin{figure}[tb]
\includegraphics[width=\columnwidth]{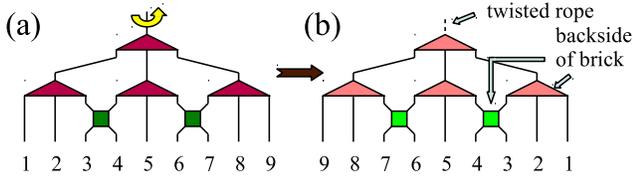}
\caption{(Color online) A real-life nine-site open boundary MERA  constructed by \textit{bricks}, representing tensors, and \textit{ropes}, representing bonds. (a) is the original structure. We can easily invert the structure by rotating the whole structure about the central rope by $180^\circ$. The top rope is twisted, represented by  dashed lines, and all the bricks turn to their backside, represented by a lighter color.}
\label{inverMERA}
\end{figure}

\begin{figure}[tb]
\includegraphics[width=0.9\columnwidth]{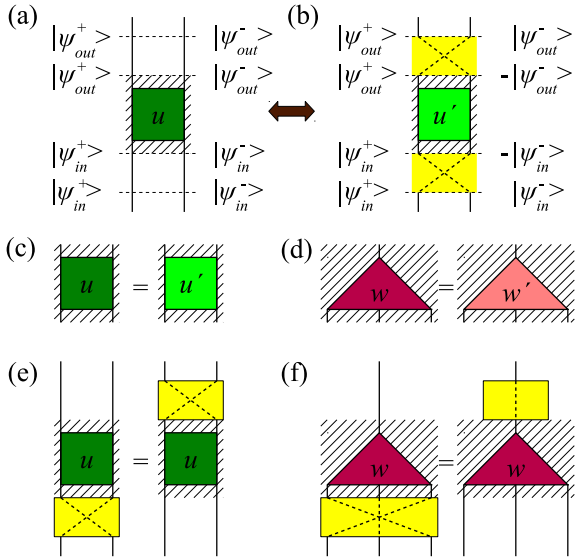}
\caption{ (Color online) (a) Operating $u$ on an inversion (anti-)symmetric state. (b)   Rotating $u$ to obtain its backside, with  bonds  swapped and twisted while holding the open ends of the  bonds fixed.  After the rotation, the originally inversion (anti-)symmetric state becomes (minus) itself, and the result after the $u'$ operation is also (minus) itself. (c) $u$ and $u^{\prime}$ are equal since the operation of $u$ and $u^{\prime}$ is the same. (Shaded area is to make it clear that what we mean by $u^{\prime}$ does not include the dashed bonds.) It's the same for $w$ as shown in (d). We further treat the yellow area (containing swapped and twisted bonds) as a local inversion operator.  In analogy of Fig.~\ref{jump_move}, we obtain the jump rules for inversion symmetric MERA as shown in (e) and (f).
}\label{inverUW}
\end{figure}

First, define a twisted \textit{rope} (bond)  as an inverted state. Note that in this definition, if  the rope is twisted twice, it is equivalent to an untwisted rope since inverting a state twice will not change the state. Also at the physical layer,  a twisted rope is equivalent to a untwisted rope since the inverted state corresponds to the state itself. For each bond, there are two types of basis states: symmetric states, denoted by $|S\rangle $,  and anti-symmetric states, $|A\rangle$. When the bond is twisted, the basis states would become $|S\rangle $ and $-|A\rangle $. 
Now, given an arbitrary $u$ (Fig.~\ref{inverUW}(a)), we can obtain the inverted $u$ by rotating the tensor while fixing the four external bonds (Fig.~\ref{inverUW}(b)). Therefore, operating $u$ on a state $|\psi_{in}\rangle$ is equivalent to taking the following sequence: (1) Invert the incoming state $|\psi_{in}^\pm \rangle$. Note the bonds are also twisted, thus the basis states are inverted (Fig.~\ref{inverMERA}).  (2) Operate $u^{\prime}$ on the inverted state, and (3)  invert the outgoing state $|\psi_{out}^\pm\rangle $. Now if the tensor is inversion symmetric, it preserves the symmetry property of the incoming state: 
\begin{equation}
|\psi_{out}^{\pm}\rangle  = u|\psi_{in}^{\pm}\rangle 
\label{u_psi}
\end{equation} 
During step (1), inverting the incoming symmetric state $|\psi_{in}^{\pm}\rangle$ gives $ \pm|\psi_{in}^{\pm}\rangle $. Since the final outgoing state is still $|\psi_{out}^{\pm}\rangle $,  so the state before the final inversion in step (3) should be $\pm|\psi_{out}^{\pm}\rangle $ (Fig.~\ref{inverUW}(b)) . Combining these two equations, we get:
\begin{equation}
u^{\prime} (\pm|\psi_{in}^{\pm}\rangle ) = \pm|\psi_{out}^{\pm}\rangle 
\label{up_psi}
\end{equation}
Comparing Eqs.~(\ref{u_psi}) and (\ref{up_psi}), finally we have (Fig.~\ref{inverUW}(c)),
\begin{equation}
u = u^{\prime},
\end{equation}
which is exactly an inversion symmetric tensor. It is easy to show that the $w$'s also have this property; therefore, our brick-and-rope representation is self-consistent.

 We can treat Fig.~\ref{inverUW}(e) and (f) as the jump rules for  inversion symmetric tensors. The power of this representation can be shown in Fig.~\ref{inverMERA}. We can either use the jump rules layer by layer to obtain Fig.~\ref{inverMERA}(b), or we can simply geometrically invert the whole structure with the above interpretation for a twisted rope. The final result is the same but the whole process becomes much simpler, and it helps us move the inversion operator to the boundary of causal cone.

\begin{figure}[tbp]
\includegraphics[width=0.9\columnwidth]{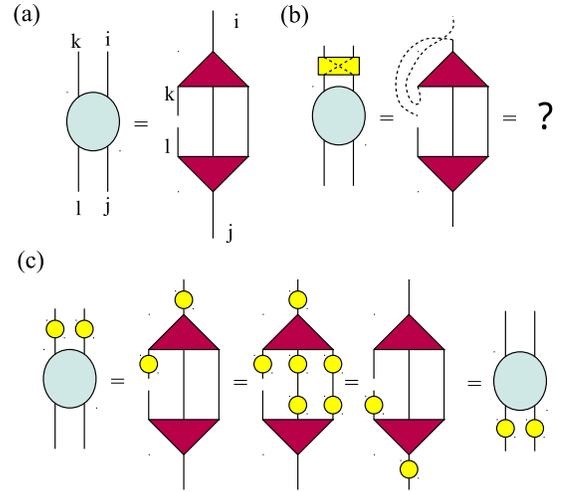}
\caption{(Color online) (a) Contraction of two bonds between inversion symmetric $w$ and $w^\dagger$. (b) The resulting tensor is generally not inversion symmetric. (c) For  internal symmetries, the resulting tensor is always symmetric.
}\label{contrac2w}
\end{figure}
\begin{figure*}[t]
\includegraphics[width=1.6\columnwidth]{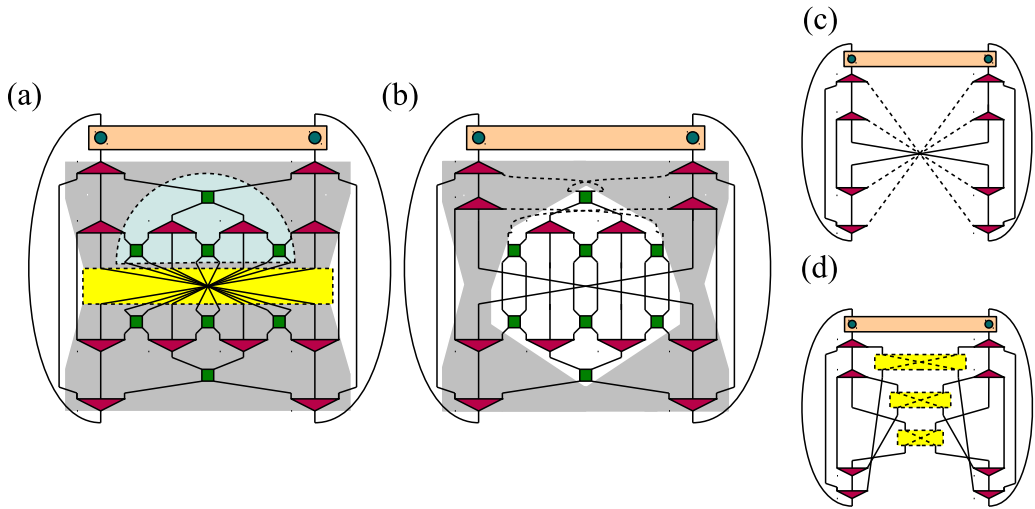}
\caption{(Color online) Reduction of the causal cone for an inversion symmetric string-order parameter. (a)  shows the original diagram with ten inverted  physical sites, and the shaded area is the original causal cone. (b) We invert the blue area (semi-circle) using the brick-and-rope representation. By doing so, the causal cone is much simplified. By contracting the tensors outside the causal cone, we obtain (c). In the sense of jump rule, the yellow box is the original inversion operator, which is lifted to the boundary of the (new) causal cone as shown in (d).
}\label{inverSO_MERA}
\end{figure*}

\begin{figure}[tbp]
\includegraphics[width=0.9\columnwidth]{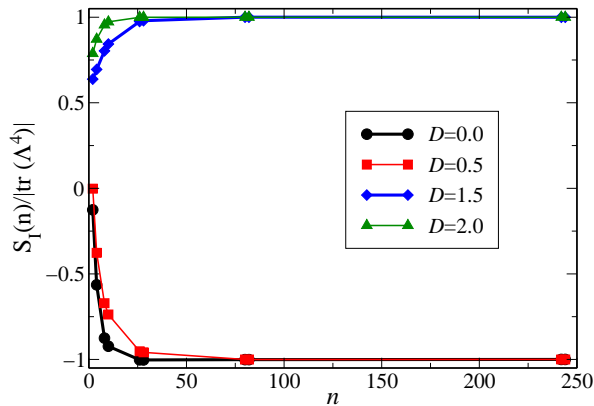}
\caption{(Color online) Inversion symmetric  string-order parameter for $N=486$ and $\chi=12$($\chi_{\rm even}=\chi_{\rm odd}=\chi/2$).
}\label{Data_ISO}
\end{figure}

Finally, we want to emphasize one key difference between the global inversion symmetry and other internal symmetries. At the first glance, the jumps rules Figs.~\ref{inverUW}(e), and (f) looks similar to Figs.~\ref{jump_move}(a), and  (b). However, taking  a closer look,  we find that the inversion operator can only jump between \textit{fixed} in- and out-bonds. On the other hand, for any internal symmetry, we have the freedom to  choose which bonds are designated as \textit{in} and \textit{out}.  In the tensor contraction process, we normally contract two tensors at one time, so the operation generally produces some intermediate tensors. For  an internal symmetry, these intermediate tensors are all symmetric. However, for the inversion symmetry, only the  \textit{final} tensor is inversion symmetric. For example, in Fig.~\ref{contrac2w}(a), we contract two bonds between two tensors $w$ and  $w^\dagger$,  which is a typical process in MERA.  The resulting tensor (Fig.~\ref{contrac2w}(b)) for inversion symmetric $w$ and $w^\dagger$ is in general not inversion symmetric, while for internal symmetries, the intermediate tensor is still symmetric (Fig.~\ref{contrac2w}(c)). On the other hand,  although the intermediate tensors are not inversion symmetric, the final tensor will be symmetric if the full diagram is inversion symmetric. In fact, most of the diagrams in MERA are not inversion symmetric; e.g., two-thirds of the environment diagrams are not inversion symmetric.\cite{MERA_algorithm} However,  linear combination of these diagrams is inversion symmetric. In practice,  this indicates that one can not use the inversion symmetry to reduce the computational cost as in the $Z_2$ case.

\section{Detecting Symmetry protected Topological Phases with MERA}
\label{sec:StringOrder}
In Ref.~\onlinecite{Pollmann2012}, Turner and Pollmann proposed nonlocal string-order parameters to detect the topological phases protected by  inversion and time-reversal symmetry. Although these order parameters are originally constructed  in the iMPS formalism, direct calculation of these parameter  by other methods should give the same results. In MERA, the calculation of these order parameters (Figs.~\ref{inverSO_MERA} and \ref{TRSO_MERA}) is generally much more involved than that of a two or three-site correlator. Using the  symmetric MERA discussed above,  we can apply  the jump rules (Figs.~\ref{jump_move} and \ref{inverUW}) to \textit{lift} the symmetry operators and  the final causal cone has the same structure as that for a multi-site  correlator, and the computation is greatly simplified. 

\subsection{ String order in the presence of inversion symmetry}
  The order parameter for the inversion symmetry can be defined as the overlap of an infinite chain with a reverted segment,
\begin{equation}
 S_I(n) = \langle \psi|I_{1,n}|\psi\rangle, 
\end{equation}
where $I_{1,n}$ is the inversion of the site $1$ to $n$ part of the chain. The limit as $n\to \infty$ is\cite{Pollmann2012} 
\begin{equation}
\lim_{n\to \infty} S_I(n)=\pm {\rm Tr} (\Lambda^4),
\end{equation}
 where the diagonal matrix $\Lambda$ contains the Schimidt values $\lambda_\alpha$ of Schimidt decomposition of the ground state wave function,
 \begin{equation}
 |\psi\rangle=\sum_{\alpha} \lambda_\alpha |L_\alpha\rangle |R_\alpha\rangle,
 \end{equation}
where $|L_\alpha\rangle$ and $|R_\alpha\rangle$ are orthonormal basis vectors of the left and right partitions, respectively.
For a topologically trivial state, the sign of $S_I(n)$ is positive, while for a topological  nontrivial state, the sign is negative.\cite{Pollmann2012}

Figure~\ref{inverSO_MERA}(a) shows  an example of  a spin chain with ten inverted sites ($n=10$). Using the brick-and-rope representation, we invert all the tensors in the blue half-circle to obtain Fig.~\ref{inverSO_MERA}(c), which can be easily calculated. If the number of inverted sites is equal to $3^k\pm1, k\in \mathbb{N}$,  as in this case, the boundary of the inversion operator is directly below the density matrix (Fig.~\ref{inverSO_MERA}(b)),  the causal cone can be greatly simplified. For all other cases, the computation is more complicated. 

For  a periodic chain with even number of sites, the ground states in  the Haldane and the large-$D$ phases are both inversion symmetry even, and we can not use the level crossing method as in Sec.~\ref{inversion_MERA} to distinguish these states. Figure~\ref{Data_ISO} shows the results of the normalized string-order parameter $S_I(n)/|{\rm tr}(\Lambda^4)|$ for several $n$'s up to $n=244$ in a chain of size $N= 486$. In this case, we take $|{\rm tr}(\Lambda^4)|=|S_I(244)|$ as it has reached the asymptotic value. It is clear that $S_I(n))/|{\rm tr}(\Lambda^4)|$ reaches to $\pm 1$ as $n$ grows. For $D>1$, the sign is positive, indicating a topologically trivial (large-$D$) phase. For $D<1$, the sign is negative, indicating a topologically nontrivial (Haldane) phase. 

\subsection{String order parameters in the presence of time-reversal symmetry}

\begin{figure}[tb]
\includegraphics[width=0.9\columnwidth,clip]{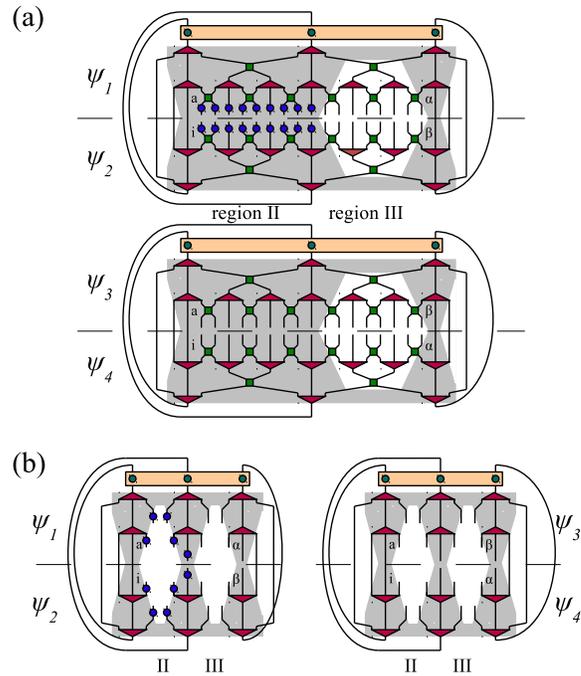}
\caption{ (Color online) Diagrammatical representation of the $n=9$ time-reversal string-order parameter in MERA. (a) The original diagram. The blue circles are the $e^{i\pi S^y}$ operators and the shaded area denotes the original causal cone. 
(b) The simplified diagram when the symmetry operators are lifted to the boundary of a new causal cone. 
}\label{TRSO_MERA}
\end{figure}
\begin{figure}[tb]
\includegraphics[width=0.9\columnwidth,clip]{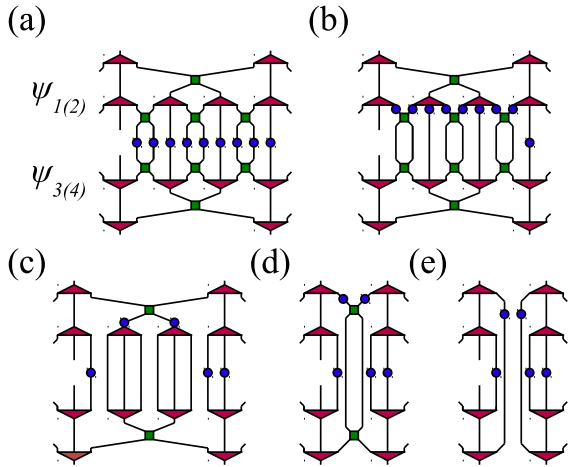}
\caption{(Color online) Reduction of the causal cone for $S_{\rm TR}(n)$.  (a) Contraction of $\psi_1$ and $\psi_3$ (or $\psi_2$ and $\psi_4$) in Region II. By using the jump rules, and  Eqs.~(\ref{u_property}) and (\ref{w_property}) successively, we obtain the intermediate diagrams (b),(c), and (d). Finally we obtain (e). For region III, the diagrams are exactly the same except there are no symmetry operators. 
}\label{TRSO_MERA_process}
\end{figure}

\begin{figure}[bt]
\includegraphics[width=0.9\columnwidth,clip]{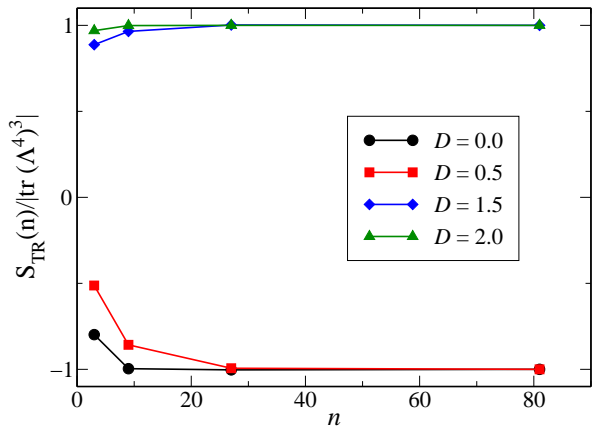}
\caption{(Color online) Time-reversal symmetric string-order parameters  for $N=486$ and $\chi=12$ ($\chi_{\rm even}=\chi_{\rm odd}=\chi/2$). }\label{Data_TRSO}
\end{figure}

For the time-reversal symmetry, the order parameter is defined as
\begin{equation}
S_{\rm TR}(n) = d^n \langle \psi_2|(|R_{1n}\rangle \langle R_{1n}|)\mbox{Swap}_{n+1,2n}|\psi_2\rangle, 
\end{equation}
where 
\[
|\psi_2\rangle  = |\psi\rangle  \otimes |\psi\rangle
\] 
is  two copies of the same wave function, 
and 
\[
|R_{1n}\rangle  = \prod_{k=1}^n \big(\frac{1}{ \sqrt{3}} \sum_{j_k} [e^{i\pi S^y}]_{j_k j_k^{\prime}} |j_k\rangle  \otimes |j_k^{\prime}\rangle 
\big)
\]
 is the TR times the complex conjugation inserted from site $1$ to site $n$. The Swap$_{n+1, 2n}$ is the swap of site $n+1$ to site $2n$ between the two copies. As $n \rightarrow \infty$, $S_{\rm TR}(n)$ reaches an asymptotic value, 
\begin{equation}
\lim_{n \to \infty}S_{\rm TR}(n) = \pm ({\rm tr }(\Lambda^4))^3.
\end{equation}
again, the sign is either  positive or negative for a  topologically trivial or non-trivial phase, respectively.

Figure~\ref{TRSO_MERA} shows the diagrammatic representation of the order parameter $S_{\rm TR}$ for $n=9$. It is more complicated as there are four wave functions with  three domain walls. Again, here we choose a convenient setup of putting the domain walls at the sites of a three-site density matrix (Fig.~\ref{TRSO_MERA}(a)).  By the application of  jump rules for $Z_2$ symmetric tensors (Fig~\ref{jump_move}), we can lift the TR symmetry operators so that the causal cone is equivalent to that of a  three-site correlator among sites at the domain walls.  All the other tensors outside this causal cone are contracted to identity (Fig.\ref{TRSO_MERA}(b)). For this choice of domain walls locations, the available  $n = 3^k, k \in \mathbb{N}$.

For  even number of sites, both the large-$D$ and the Haldane phases are TR symmetry even, and we can not use level crossing to distinguish these phases. The string-order parameter, on the other hand, can still be used to distinguish these phases. Figure~\ref{Data_TRSO} shows  $S_{\rm TR}(n)$ for several $n$ in a chain of $N=486$.    As $n$ grows, $S_{TR}(n)$ reaches  an asymptotic value. The sign of $S_{\rm TR}(n)$  is either positive  or negative for the Haldane phase or the large-$D$ phase as expected.

\section{Symmetry Protected Renormalization Group Flow}

The extra dimension of MERA  provides information about the RG flow during the coarse-graining transformation at different length scales. At the $\tau$-th layer, we obtain an effective Hamiltonian $H^\tau$ and the coarse-grained wave function $|\psi^\tau\rangle$. If symmetric disentanglers and isometries are used in the entanglement renormalization process, we could obtain information about the  symmetry-protected RG flow. In principle, one can take the symmetric disentanglers $u$ and isometries $w$ from the finite-size MERA for the buffer layers and construct a scale invariant MERA\cite{Pfeifer:2009ve} to obtain the fixed point tensors for the SPT phases.\cite{Singh:2013ly,Huang:2013uq} 
Here we take a  slightly different approach. We calculate the inversion symmetric order parameter at each layer as the system is coarse-grained.  We expect the state will flow to either the AKLT state for $D<D_c$ or the large-$D$ state for $D>D_c$ in the presence of symmetry.  Therefore, the asymptotic value of the order parameter should flow to $-1/2$ for $D <D_c$ because the AKLT state has the Schimidt matrix\cite{Pollmann2012}     
\[
\Lambda_{\rm AKLT}=\left(\begin{array}{cc}
\frac{1}{\sqrt{2}} & 0 \\
0 & \frac{1}{\sqrt{2} }
\end{array}\right),
\] 
and $-{\rm Tr}(\Lambda_{\rm AKLT}^4)=-1/2$.
On the other hand, for $D>D_c$, the state is expected to flow to  a product state with $\Lambda_{\rm prod}=1$, and ${\rm Tr}(\Lambda_{\rm prod}^4)=1$.   
Figure~\ref{Data_RG} shows the order parameter for the inversion symmetry, $S_I(n)$, at different length scales  with $N=1458$ and $\chi=12$($\chi_{\rm even}=\chi_{\rm odd}=\chi/2$). The number of inverted sites for each layer is $n$=730, 244, 82, and 28, respectively. As we proceed the coarse-graining process from the physical (0$^{\rm th}$) layer to higher layers, we observe that for $D\lesssim1$, the order parameter flows to the asymptotic value $-1/2$ as expected, and flows to $1$ for large $D$. Notice in the regime of $D\gtrsim1$, the flow to the product state is much slower, indicating the ground state is far away from the fixed point,  and it requires  more RG steps (or larger $\chi$) to reach the asymptotic value $1$. This also explains why $S_I(n)$ fails to reach the asymptotic value for $1 \lesssim D <1.4$ (shaded area in Fig.~\ref{Data_RG}).   
\begin{figure}[bt]
\includegraphics[width=0.9\columnwidth,clip]{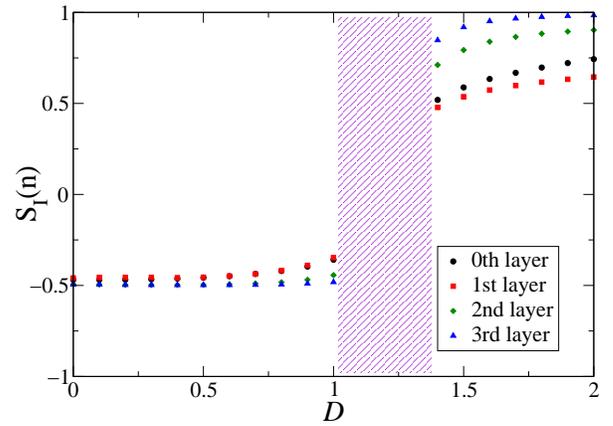}
\caption{(Color online) Order parameter for the inversion symmetry $S_I(n)$ at different length scales for a spin chain with $N=1458$ and $\chi=12$ ($\chi_{\rm even}=\chi_{\rm odd}=\chi/2$). The number of inverted sites for each layer is $n=730, 244,82$, and 28, respectively. Shaded area indicates where $S_I(n)$ fails to reach asymptotic. }\label{Data_RG}
\end{figure}

\section{Conclusion}
\label{sec:Conclusion}
Global symmetries have been used extensively 
in MERA algorithms to reduce computation efforts, and  to access different quantum number sectors.\cite{Singh:2010zr,Singh:2011uq,Singh:2012kx} 
In this paper, we demonstrate how to use symmetric tensors in MERA to study the SPT phase for a given symmetric Hamiltonian.\cite{Singh:2013ly}
We have proposed a brick-and-rope representation for an inversion symmetric MERA, which gives a simple geometrical interpretation for inversion symmetric tensors. Using the time-reversal and inversion symmetric MERA, we demonstrate that the Haldane phase in spin-1 chain is protected by these symmetries. Finally, by computing the inversion symmetric string-order parameter at different length scales, we show that entanglement renormalization with symmetric tensors indeed provides information about the symmetry-protected RG flow, and the system does flow to a RG fixed point corresponding to the SPT phase.  With the current scheme, it is easy to extend to  spin chains and ladders with higher integer spins. Also, it can also be generalized to study the fractionalization of  quasi-particles (edge states) by implementing the projective representation of the global symmetry in the symmetric tensors.

\acknowledgements
We thank G. Vidal and F. Pollmann for useful discussions. This work is partially supported by NSC in Taiwan through Grant No.
100-2112-M-002-013-MY3, 100-2923-M-004 -002 -MY3- and by NTU Grant numbers 101R891004. Travel support from NCTS in Taiwan is also acknowledged.

\begin{figure}[t]
\includegraphics[width=0.9\columnwidth]{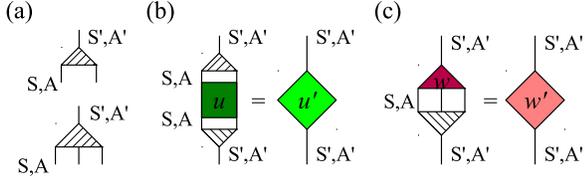}
\caption{(Color online) (a) The basis transformation tensors for $u$ (top) and $w$ (bottom), which transforms the original $\tau$th-layer symmetric basis ($S,A$) into a $(\tau+1)$th-layer symmetric basis ($S^{\prime},A^{\prime}$). Operating these tensors on (b) $u$(green) and (c) $w$(red) makes them block-diagonal in the symmetric basis ($S', A'$).
}\label{basis_transform}
\end{figure}

\appendix

\section{Fusion rules for inversion symmetry}

  In this appendix, we present an example of the inversion fusion rules for $u$ and $w$ in MERA. As in the ${Z}_2$ case, we need to construct an inversion symmetric basis at each layer of MERA. For simplicity, we take  $S=1/2$ as an example. At each physical site, there are two physical degrees of freedom, denoted by $|\uparrow\rangle$ and $|\downarrow\rangle$.  At the physical (0th) layer, the fusion rules for $u$ and $w$ are given in Tabs.~\ref{tb:Ufusion0} and \ref{tb:Wfusion0}. In terms of these basis states,  $u$ and $w$ are block-diagonal (Fig.~\ref{basis_transform}). At the first layer, for the illustration purpose, we  keep only one symmetric state, denoted by $S$, and one anti-symmetric basis state, denoted by $A$. These states can be a linear combination of the original basis states. We symmetrize and anti-symmetrize these states to obtain a new basis for the $(\tau+1)$th-layer (Tabs.~\ref{tb:Ufusion} and \ref{tb:Wfusion}. Note that we have states that are by themselves anti-symmetric (e.g. $SAS$).  Also, the symmetrized  state for  $SSA$ is  $(SSA - ASS)/\sqrt{2}$ while the anti-symmetrized state is $(SSA + ASS)/\sqrt{2}$. The reason for this can be easily observed from Fig.~\ref{inver2w} that to invert a state, we not only swap the location of the bonds, but we twist these bonds as well, which will invert the tensors below. Notice that the fusion rules in Tabs.~\ref{tb:Ufusion0} and \ref{tb:Wfusion0} only apply to the case where each site at the physical layer corresponds to one physical site. In this case, the inversion of single-site state at the physical layer is the state itself. If the sites at the physical layer are the composite of multiple physical sites, then the fusion rules in Tabs.~\ref{tb:Ufusion} and \ref{tb:Wfusion} should be used at the physical layer also. One can easily generalize the tables for a larger $\chi$, where we have $S_i, i=1, \ldots, {\chi_s}$ and $A_i,  i=1,\ldots,{\chi_a}$, with $\chi_s$ and $\chi_a$ being the bond dimensions for the symmetric and anti-symmetric sectors. 
\begin{table}[p]
\caption{Fusion rules for $u$ at the physical layer
}\label{tb:Ufusion0}
\begin{tabular}{c||c|c}
Original basis & Symmetrized($S$) & Anti-symmetrized($A$)\\
\hline \hline
$\uparrow\uparrow$ & $\uparrow\uparrow$ & -\\
$\uparrow\downarrow$, $\downarrow\uparrow$ & $(\uparrow\downarrow + \downarrow\uparrow)/\sqrt{2}$ & $(\uparrow\downarrow - \downarrow\uparrow)/\sqrt{2}$\\
$\downarrow\downarrow$ & $\downarrow\downarrow$ & -\\
\end{tabular}
\end{table}

\begin{table}[p]
\caption{Fusion rules for $w$ at the physical layer
}\label{tb:Wfusion0}
\begin{tabular}{c||c|c}
Original basis & Symmetrized($S$) & Anti-symmetrized($A$)\\
\hline \hline
$\uparrow\uparrow\uparrow$ & $\uparrow\uparrow\uparrow$ & -\\
$\uparrow\uparrow\downarrow$, $\downarrow\uparrow\uparrow$ & $(\uparrow\uparrow\downarrow + \downarrow\uparrow\uparrow)/\sqrt{2}$ & $(\uparrow\uparrow\downarrow - \downarrow\uparrow\uparrow)/\sqrt{2}$\\
$\uparrow\downarrow\uparrow$ & $\uparrow\downarrow\uparrow$ & -\\
$\uparrow\downarrow\downarrow$, $\downarrow\downarrow\uparrow$ & $(\uparrow\downarrow\downarrow + \downarrow\downarrow\uparrow)/\sqrt{2}$ & $(\uparrow\downarrow\downarrow - \downarrow\downarrow\uparrow)/\sqrt{2}$\\
$\downarrow\uparrow\downarrow$ & $\downarrow\uparrow\downarrow$ & -\\
$\downarrow\downarrow\downarrow$ & $\downarrow\downarrow\downarrow$ & -\\
\end{tabular}
\end{table}

\begin{table}[p]
\caption{Fusion rules for $u$ at upper layers
}\label{tb:Ufusion}
\begin{tabular}{c||c|c}
Original basis & Symmetrized($S^{\prime}$) & Anti-symmetrized($A^{\prime}$)\\
\hline \hline
$SS$ & $SS$ & -\\
$SA$, $AS$ & $(SA - AS)/\sqrt{2}$ & $(SA + AS)/\sqrt{2}$\\
$AA$ & $AA$ & -\\
\end{tabular}
\end{table}

\begin{table}[p]
\caption{Fusion rules for $w$ at upper layers
}\label{tb:Wfusion}
\begin{tabular}{c||c|c}
Original basis & Symmetrized($S^{\prime}$) & Anti-symmetrized($A^{\prime}$)\\
\hline \hline
$SSS$ & $SSS$ & -\\
$SSA,ASS$ & $(SSA - ASS)/\sqrt{2}$ & $(SSA + ASS)/\sqrt{2}$\\
$SAS$ & - & $SAS$\\
$SAA,AAS$ & $(SAA + AAS)/\sqrt{2}$ & $(SAA - AAS)/\sqrt{2}$\\
$ASA$ & $ASA$ & -\\
$AAA$ & - & $AAA$\\
\end{tabular}
\end{table}
\bibliography{Topological_Phase_MERA}

\end{document}